\documentclass{scrartcl}

\usepackage{amssymb,amsmath,amsthm}
\usepackage[numbers]{natbib}
\usepackage{xspace}
\usepackage{enumerate}

\DeclareMathAlphabet{\mathpzc}{OT1}{pzc}{m}{it}

\begin{document}

% THEOREMS -------------------------------------------------------
\newtheorem{thm}{Theorem}[section]
\newtheorem{cor}[thm]{Corollary}
\newtheorem{lem}[thm]{Lemma}
\newtheorem{prop}[thm]{Proposition}
\theoremstyle{definition}
\newtheorem{defn}[thm]{Definition}
\newtheorem{cond}[thm]{Condition}
\newtheorem{ass}[thm]{Assumption}
\theoremstyle{remark}
\newtheorem{rem}[thm]{Remark}
\newtheorem{example}[thm]{Example}
%\numberwithin{equation}{section}

\newcommand{\norm}[1]{\left\Vert#1\right\Vert}
\newcommand{\abs}[1]{\left\vert#1\right\vert}
\newcommand{\set}[1]{\left\{#1\right\}}
\newcommand{\Ind}[1]{\mathbf{1}_{\left\{#1\right\}}}
\newcommand{\ii}{{I \setminus \{i\}}}
\newcommand{\RR}{\mathbb{R}}
\newcommand{\CC}{\mathbb{C}}
\newcommand{\QQ}{\mathbb{Q}}
\newcommand{\PP}{\mathbb{P}}
\newcommand{\FF}{\mathbb{F}}
\newcommand{\NN}{\mathbb{N}}
\newcommand{\EE}{\mathbb{E}}
\newcommand{\cF}{\mathcal{F}}
\newcommand{\cC}{\mathcal{C}}
\newcommand{\cU}{\mathcal{U}}
\newcommand{\cA}{\mathcal{A}}
\newcommand{\cI}{\mathcal{I}}
\newcommand{\cD}{\mathcal{D}}
\newcommand{\cR}{\mathcal{R}}
\newcommand{\St}{(S_t)_{t \ge 0}}
\newcommand{\Xt}{(X_t)_{t \ge 0}}
\newcommand{\Vt}{(V_t)_{t \ge 0}}
\newcommand{\Rminus}{\mathbb{R}_{\leqslant 0}}
\newcommand{\pd}[2]{\frac{\partial #1}{\partial #2}}
\newcommand{\Rplus}{\mathbb{R}_{\geqslant 0}}
\newcommand{\pdn}[3]{\frac{\partial^{#1} #2}{\partial #3^{#1}}}
\newcommand{\scal}[2]{\left\langle{#1},{#2}\right\rangle}
\newcommand{\supp}{\operatorname{supp}}
\newcommand{\rank}{\operatorname{rank}}
\newcommand{\relint}{\operatorname{ri}}  % relative interior of convex set
\newcommand{\Int}{\operatorname{int}}  % relative interior of convex set
\newcommand{\aff}{\operatorname{aff}}    % affine hull
\newcommand{\dom}{\operatorname{dom}}    % effective domain
\newcommand{\bd}{\operatorname{bd}}    % relative boundary
\newcommand{\rbd}{\operatorname{rbd}}    % relative boundary
\newcommand{\cl}{\operatorname{cl}}    % relative boundary
\newcommand{\E}[1]{\mathbb{E}\left[#1\right]}
\newcommand{\Ex}[2]{\mathbb{E}^{#1}\left[#2\right]}
\newcommand{\Econd}[2]{\mathbb{E}\left[\left.#1\right|#2\right]}
\newcommand{\Excond}[3]{\mathbb{E}^{#1}\left[\left.#2\right|#3\right]}
\newcommand{\imp}{\text{imp}}
\newcommand{\loc}{\text{loc}}
\newcommand{\BS}{\text{BS}}
\newcommand{\osigma}{\overline{\sigma}}

\def\lsnc{\ensuremath{\mathrm{LSNC-}\chi^2}}
\def\nc{\ensuremath{\mathrm{NC-}\chi^2}}
%% ================================================================ %%
%% ================================================================ %%

\title{A remark on Gatheral's `most-likely path approximation' of implied volatility}

\author{Martin Keller-Ressel and Josef Teichmann}
%\email{\{kemartin\}@math.ethz.ch} \keywords{}
\date{September 5th, 2014}
\maketitle %\pagestyle{myheadings} \frenchspacing
\begin{abstract}
We give a new proof of the representation of implied volatility as a
time-average of weighted expectations of local or stochastic volatility. With
this proof we clarify the question of existence of `forward implied variance' in the original
derivation of Gatheral, who introduced this representation
in his book `The Volatility Surface'. 
\end{abstract}

\section{Gatheral's most-likely path approximation}
In his book `The Volatility Surface -- A Practitioners Guide', Jim Gatheral presents an approximation formula for the implied volatility of a European option, when the underlying stock follows a general diffusion process
\begin{equation}\label{Eq:loc_vol}
\frac{dS_t}{S_t} = \mu(t,S_t)\,dt + \sigma(t,S_t)\,dW_t\;.
\end{equation}
The `most-likely path approximation' to implied Black-Scholes volatility in this model consists of two parts: The first part is the assertion that implied variance -- the square of implied volatility -- can be written as a time-average of weighted expectations of $\sigma^2(t,S_t)$:  
\begin{equation}\label{Eq:imp_vol_exact}
\sigma^2_\imp(K,T) = \frac{1}{T}\int_0^T{\Ex{\mathbb{G}_t}{\sigma^2(t,S_t)}dt}\;.
\end{equation}
Here, the measures $\mathbb{G}_t$ are given by their Radon-Nikodym derivatives with respect to the risk-neutral measure $\QQ$,
\begin{equation}\label{Eq:Radon_nikodym}
\frac{d\mathbb{G}_t}{d\QQ} = \frac{S_t^2 \Gamma_\BS(S_t,\osigma_{K,T}(t))}{\E{S_t^2 \Gamma_\BS(S_t,\osigma_{K,T}(t))}},
\end{equation}
where $\osigma_{K,T}(t)$ is a function that is yet to be specified, $\Gamma_\BS$ denotes the Black-Scholes Gamma and expectations are always taken to be under the risk-neutral pricing measure. Let us emphasize that \eqref{Eq:imp_vol_exact} is an exact formula, and that it is the second part of the method where the approximation happens: Gatheral argues that the density \eqref{Eq:Radon_nikodym} is concentrated (as a function of $(t,S)$) close to a narrow ridge connecting today's stock price $S_0$ to the strike price $K$ at time $T$, and claims that a good approximation to \eqref{Eq:imp_vol_exact} is to evaluate it as if the density was \emph{entirely concentrated} on this ridge\footnote{See \citet[Page~29ff]{Gatheral2006} for details.}. In the terminology of Gatheral this ridge is called the most-likely path and the described approximation method the most-likely path approximation. 
Extensions of the representation \eqref{Eq:Radon_nikodym} have been proposed e.g. by \citet{Guyon2013} for implied correlations.\\

In this note we will only be concerned with the first part of Gatheral's method, i.e. the derivation of the exact equation \eqref{Eq:imp_vol_exact}, and in particular the definition of the yet unknown function $\osigma_{K,T}(t)$. \citet{Gatheral2006} defines on page 27 first the `Black-Scholes forward implied variance' $v_{K,T}(t)$ by
\begin{equation}\label{Eq:vKT_wrong_def}
v_{K,T}(t) = \frac{\E{\sigma^2(t,S_t) S_t^2 \Gamma_\BS(S_t,\osigma_{K,T}(t))}}{\E{S_t^2 \Gamma_\BS(S_t,\osigma_{K,T}(t))}},
\end{equation}
and then, in the equation below, the quantity $\osigma_{K,T}(t)$ by
\begin{equation}\label{Eq:sigma_from_v}
\osigma_{K,T}^2(t) = \frac{1}{T-t}\int_t^T{v_{K,T}(u)du}\;.
\end{equation}
Differentiating \eqref{Eq:sigma_from_v} and inserting into \eqref{Eq:vKT_wrong_def} yields an ordinary differential equation for $\osigma_{K,T}(t)$. This definition through an ODE leaves open the question whether (and under which conditions) the quantities $v_{K,T}(t)$ and $\osigma_{K,T}(t)$ actually exist\footnote{See also \citet[Sec.~2.3]{Lee2004a}, who remarks that the proof in \citet{Gatheral2006} hinges upon the assumption of the existence of $v_{K,T}(t)$.}. We will show that a simpler definition of $\osigma_{K,T}(t)$ can be given, which clarifies the problem of existence, implies equation \eqref{Eq:vKT_wrong_def} and \eqref{Eq:sigma_from_v} and finally leads to a proof of the implied volatility representation \eqref{Eq:imp_vol_exact}.

\section{A new proof of the implied volatility representation}

For our proof of the implied volatility representation we assume that the stock price follows an It\^o-process with respect to the risk-neutral measure $ \QQ $ (with respect to which all expectations are taken) of the form
\begin{equation}\label{Eq:loc_vol}
\frac{dS_t}{S_t} = r\,dt + \sigma_t\,dW_t\; ,
\end{equation}
such that the discounted stock price ${(e^{-rt}S_t)}_{0 \leq t \leq T}$ is a square-integrable martingale. The volatility process $ \sigma $ is a general predictable, $W$-integrable process. This setup covers in particular local volatility models, where $\sigma_t = \sigma(t,S_t)$ and stochastic volatility models where $\sigma_t = \sigma(t,V_t)$ and $V_t$ is a stochastic factor driving the volatility. We fix a terminal time $T$ and assume that $S$ is non-deterministic in the sense that $\PP(S_t \neq S_T) > 0$ for all $t \in [0,T]$. Fixing also a strike price $K$ we are ultimately interested in the implied Black-Scholes volatility $\sigma_\text{imp}(T,K)$ for a European option with expiry $T$ and strike $K$ in the above model. 

\subsection{A regime-switching model and implied forward total variance}

To start our derivation, we associate for each $u \in [0,T]$ and $\Sigma^u \ge 0$ the `regime-switching' process $S^u$ to $S$, given by
\begin{equation}\label{Eq:Stau}
\begin{split}
\frac{dS^u_t}{S^u_t} &= r\,dt + \sigma_t\,dW_t\quad t \in [0,u] \\
\frac{dS^u_t}{S^u_t} &= r\,dt + \Sigma^u \,dW_t\quad t \in [u,T]. 
\end{split}
\end{equation}
The process $S^u$ switches, at time $t = u$, from the dynamics \eqref{Eq:loc_vol} to Black-Scholes dynamics with constant volatility $\Sigma^u$. It should be obvious, that $S^T = S$, while $S^0$ is simply a Black-Scholes model with volatility $\Sigma^0$. In what follows, it will be helpful to consider the \emph{total variance} $w_u = (T-u)(\Sigma^u)^2$ instead of $\Sigma^u$. By simple conditioning, the price of a put option on $S^u$ with strike $K$ and maturity $T$ is given by 
\[e^{-rT} \E{(K - S_u)_+} = e^{-ru} \E{e^{-r(T-u)}\Econd{(K - S_u)_+}{\cF_u}} = e^{-ru}\E{P_\text{BS}(u,S_u,T,K;w_u)},\]
where $P_\text{BS}(u,S,T,K;w)$ is the Black-Scholes put-price parametrized by total variance, i.e.
\[ P_\text{BS}(u,S,T,K;w) = e^{-r(T-u)}K\Phi(-d_2) - S \Phi(-d_1) \]
and
\[d_{1,2}(w) = \frac{\log\left(\frac{e^{r(T-u)}S}{K}\right)}{\sqrt{w}} \pm \frac{\sqrt{w}}{2}.\]
\begin{defn}\label{Def:wu}
For $u \in [0,T)$ we define the \textbf{implied forward total variance} $\hat w_u = \hat w_u(T,K) \ge 0$ as the solution of
\begin{equation}
e^{-ru}\E{P_\text{BS}(u,S_u,T,K;\hat w_u)} = e^{-rT} \E{(K - S_T)_+}
\end{equation}
i.e. $\hat w_u$ is the total variance $w_u = (T-u) (\Sigma^u)^2$ that has to be chosen in the regime-switching model \eqref{Eq:Stau} such that the resulting put-price coincides with the put-price from the original model \eqref{Eq:loc_vol}.
\end{defn}

\begin{prop}
There exists a unique positive deterministic function $u \mapsto \hat w_u$, such that the equality
\begin{equation}\label{Eq:w}
e^{-ru}\E{P_\text{BS}(u,S_u,T,K;\hat w_u)} = e^{-rT} \E{(K - S_T)_+}
\end{equation}
is satisfied for all $u \in [0,T]$. 
\end{prop}
\begin{proof}
For $w = 0$, the Black-Scholes price $e^{-ru}P_\text{BS}(u,S_u,K,T;w)$ equals $e^{-ru}(e^{-r(T-u)}K - S_u)_+$. Since ${(e^{-ru}S_u)}_{0 \leq u \leq T}$ is a martingale, we have by Jensen's inequality that
\[\
e^{-ru}\E{P_\text{BS}(u,S_u,K,T;0)} = e^{-ru}\E{(e^{-r(T-u)}K - S_u)_+} \le e^{-rT}\E{(K - S_T)_+} \;.
\]
For $w \to \infty$ the Black-Scholes price $P_\text{BS}(u,S_u,K,T;w)$ approaches $e^{-r(T-u)}K$. In this case we get
\[
e^{-ru}\E{P_\text{BS}(u,S_u,T,K;\infty)} = e^{-rT}K \ge e^{-rT}\E{(K-S_T)_+} \;.
\]
In addition $w \mapsto P_\BS(t,S_t,T,K;w)$ is for any given $S_t$ a continuous and strictly monotone increasing function (here we need the non-degeneracy assumption on $S$), hence also $w \mapsto \E{P_\BS(t,S_t,T,K;w)}$ is. Therefore we conclude that \eqref{Eq:w} has a unique solution $\hat w_u$ for each $u \in [0,T]$. 
\end{proof}
\begin{rem}
Notice that the previous proof holds in fact for semi-martingales $ S $, such that $ {(\exp(-rt)S_t)}_{0 \leq t \leq T} $ is a martingale, so neither square integrability nor absence of jumps are needed. However, we do not get regularity assertions for $ u \mapsto \hat w_u $.
\end{rem}

\subsection{Main Result}
We now present our main result on the implied forward total variance $\hat w_u$. Here the assumption of continuous trajectories is really needed, as well as the following $L^2$-continuity assumption:
\begin{ass}\label{Ass:msq}
We assume that $\sigma_u$ is mean-square continuous, i.e. the map $ [0,T] \ni u \mapsto  \sigma^2_u \in L^2(\Omega,\QQ)$ is continuous with respect to the $L^2$-topology.
\end{ass}

\begin{thm}
Under assumption~\ref{Ass:msq} the mapping $u \mapsto \hat w_u$ is in $C^1[0,T) \cap C^0[0,T]$ and satisfies the ODE
\begin{equation}\label{Eq:TV_ODE}
\pd{\hat w_u}{u} = - \frac{\E{\phi(d_2(\hat w_u)) \sigma^2_u}}{\E{\phi(d_2(\hat w_u))}}, \qquad u \in [0,T),
\end{equation}
with terminal condition $\lim_{u \to T} \hat w_T = 0$ and where $\phi$ denotes the standard normal density. For $u = 0$ it holds that
\[\hat w_0(T,K) = T \sigma^2_\text{imp}(T,K),\]
where $\sigma_\text{imp}(T,K)$ is the implied Black-Scholes volatility for time-to-maturity $T$ and strike $K$ in \eqref{Eq:loc_vol}.
\end{thm}
\begin{rem}
Equation \eqref{Eq:TV_ODE} can be rewritten as \eqref{Eq:imp_vol_exact}. Alternatively, it can be written as
\begin{equation*}
- \pd{}{u}\hat w_u = \E{\sigma^2_u}  + \textbf{Cov} \left(\frac{\phi(d_2(\hat w_u))}{\E{\phi(d_2(\hat w_u))}},  \sigma^2_u \right), 
\end{equation*}
i.e., the rate of decrease in total implied variance is given by expected instantaneous stochastic volatility plus a correction term that accounts for correlation effects between $\sigma_u$ and $S_u$ in a highly non-linear way. 
\end{rem}

\begin{proof}
We set
\[F(u,w)= e^{-ru}\E{P_\text{BS}(u,S_u,T,K;w)}.\]
Note that the derivative of $P_\text{BS}$ with respect to total variance $w$ is given by
\[\pd{}{w} P_\text{BS}(u,S,T,K;w) = \frac{1}{2\sqrt{w}} S \phi(d_1),\]
which, inserting $S = S_u$,  is uniformly integrable in $w$ on each interval $(\epsilon,\infty)$, $\epsilon > 0$. Hence for $w \in (0,\infty)$,
\begin{equation}\label{eq:dw}
\pd{}{w} F(u,w)= \frac{e^{-ru}}{2\sqrt{w}}\E{S_u \phi(d_1(w))} = \frac{e^{-rT}}{2\sqrt{w}}\E{\phi(d_2(w))}.
\end{equation}
Applying Ito's formula and using the martingale property of $S$ we obtain
\begin{equation}\label{eq:after_ito}
\pd{}{u} F(u,w) = e^{-ru} \E{-r P_\text{BS} + \pd{}{u}P_\text{BS} + \pd{}{S} P_\text{BS} r S_u + \frac{1}{2} \pdn{2}{}{S} P_\text{BS} S_u^2 \sigma^2_u}.
\end{equation}
Parameterized by total implied variance, the Black-Scholes put-price $P_\text{BS}$ satisfies 
\[
-r P_\text{BS} + \pd{}{u} P_\text{BS} + r S \pd{}{S} P_\text{BS}  = 0 \; ,
\]
such that \eqref{eq:after_ito} simplifies to
\begin{equation}\label{eq:du}
\pd{}{u} F(u,w) = e^{-ru} \frac{1}{2} \E{ \pdn{2}{}{S}P_\text{BS} S_u^2 \sigma^2_u} = \frac{1}{2} \frac{e^{-rT}K}{\sqrt{w}} \E{\phi(d_2(w)) \sigma^2_u}.
\end{equation}
Note that due to Assumption~\ref{Ass:msq} both $\partial_u F(u,w)$ and $\partial_w F(u,w)$ are continuous. Furthermore, recall that $\hat w_u$ is given in Definition~\ref{Def:wu} by the implicit equation
\begin{equation}\label{Eq:implicit}
F(u,\hat w_u) = e^{-rT} \E{(K - S_T)_+}, 
\end{equation}
where the right hand side depends neither on $u$ nor on $\hat w_u$. Let us first examine the boundary behavior of $F(u,w)$. We easily derive that
\begin{align*}
\lim_{w \to 0} F(u,w) &= \E{\left(e^{-rT}K - e^{-ru}S_u\right)_+} ,\\
\lim_{w \to \infty} F(u,w) &= e^{-rT} K ,\\
\lim_{u \to 0} F(u,w) &= P_\text{BS} (0,S_0,K;w),\\
\lim_{u \to T} F(u,w) &= e^{-rT} \E{\Phi(-d_2(w))K - \Phi(-d_1(w))S_T}.
\end{align*}
By Jensen's inequality and the assumptions on the non-degeneracy of $S$ it holds that 
\[\E{\left(e^{-rT}K - e^{-ru}S_u\right)_+} < e^{-rT}\E{(K-S_T)_+} < e^{-rT} K\]
for all $u \in [0,T)$. From \eqref{eq:dw} we see that $\partial_w F(u,w) > 0$ and hence $w \mapsto F(u,w)$ is increasing for $w \in (0,\infty)$. 
Altogether, it follows that for each $u \in [0,T]$ a unique $\hat w_u$ solving \eqref{Eq:implicit} exists. In addition, by the implicit function theorem, $\hat w_u$ is in $C^1[0,T) \cap C^0[0,T]$ with derivative 
\[\pd{}{u} \hat w_u = -\frac{\partial_u F(u,w)}{\partial_w F(u,w)} = - \frac{\E{\phi(d_2(w_u))\sigma^2_u}}{\E{\phi(d_2(w_u))}},\]
where we have combined \eqref{eq:dw} and \eqref{eq:du}. The initial and terminal conditions for $\hat w_u$ at $u = 0$ and $u = T$ can be derived from the above boundary conditions for $F(u,w)$. Indeed,
\[P_\text{BS} (0,S_0,K;\hat w_0) = C(K,T)\]
implies that $\hat w_0 = T \sigma_\text{imp}^2$, where $\sigma_\text{imp}$ is the Black-Scholes implied volatility corresponding to the put-price $P(K,T)$. Finally
\[\E{\Phi(-d_2(w))K - \Phi(-d_1(w)) S_T} = P(K,T) = \E{(K - S_T)_+}\]
implies that $w = 0$ and hence both boundary conditions for $\hat w_u$ follow. 
\end{proof}

\bibliographystyle{plainnat}
\bibliography{references}

\begin{thebibliography}{3}
\providecommand{\natexlab}[1]{#1}
\providecommand{\url}[1]{\texttt{#1}}
\expandafter\ifx\csname urlstyle\endcsname\relax
  \providecommand{\doi}[1]{doi: #1}\else
  \providecommand{\doi}{doi: \begingroup \urlstyle{rm}\Url}\fi

\bibitem[Gatheral(2006)]{Gatheral2006}
Jim Gatheral.
\newblock \emph{The Volatilty Surface}.
\newblock Wiley Finance, 2006.

\bibitem[Guyon and Henry-Labord{\`e}re(2013)]{Guyon2013}
Julien Guyon and Pierre Henry-Labord{\`e}re.
\newblock \emph{Nonlinear Option Pricing}.
\newblock CRC Press, 2013.

\bibitem[Lee(2004)]{Lee2004a}
Roger Lee.
\newblock Implied volatility: Statics, dynamics, and probabilistic
  interpretation.
\newblock In \emph{Recent Advances in Applied Probability}. Springer, 2004.

\end{thebibliography}

\end{document}